\documentclass[iop,revtex4,apjl]{emulateapj}
\usepackage{graphicx,color}
\usepackage{amssymb}
\usepackage{amsmath}
\usepackage{threeparttable}
\usepackage{lscape}
\renewcommand{\baselinestretch}{1.0} 
\mathchardef\mhyphen="2D

\newcommand{\oiii}{[O\,{\sc iii}]}
\newcommand{\ovi}{O\,{\sc vi}}

\newcommand{\feii}{Fe\,{\sc ii}}

\newcommand{\siv}{S\,{\sc iv}} 
\newcommand{\sxvi}{S\,{\sc xvi}}
\newcommand{\siiv}{Si\,{\sc iv}}

\newcommand{\ciii}{C\,{\sc iii}}
\newcommand{\civ}{C\,{\sc iv}}

\newcommand{\angstrom}{\text{ \normalfont\AA}}

\mathchardef\mhyphen="2D

\definecolor{red}{rgb}{0.75,0.0,0.0}
\definecolor{yel}{rgb}{0.65,0.65,0.0}
\definecolor{grn}{rgb}{0.0,0.75,0.0}
\definecolor{blu}{rgb}{0.0,0.0,0.75}
\definecolor{gry}{rgb}{0.75,0.75,0.75}
\def\Nion{\ifmmode N_\mathrm{\scriptstyle ion} \else $N_\mathrm{\scriptstyle ion}$\fi}

\def\hi{H\,{\sc i}}

\def\ciii{C\,{\sc iii}}
\def\civ{C\,{\sc iv}}
\def\niii{N\,{\sc iii}}
\def\niv{N\,{\sc iv}}
\def\nv{N\,{\sc v}}

\def\oiii{O\,{\sc iii}}
\def\oiv{O\,{\sc iv}}
\def\ov{O\,{\sc v}}
\def\ovi{O\,{\sc vi}}

\def\neiv{Ne\,{\sc iv}}
\def\nev{Ne\,{\sc v}}
\def\nevi{Ne\,{\sc vi}}
\def\neviii{Ne\,{\sc viii}}

\def\nex{Ne\,{\sc x}}
\def\naix{Na\,{\sc ix}}

\def\mgx{Mg\,{\sc x}}
\def\mgxii{Mg\,{\sc xii}}

\def\alxi{Al\,{\sc xi}}
\def\siiv{Si\,{\sc iv}}
\def\Siii{Si\,{\sc ii}}

\def\sixii{Si\,{\sc xii}}
\def\sixiii{Si\,{\sc xiii}}
\def\sixiv{Si\,{\sc xiv}}

\def\siv{S\,{\sc iv}}

\def\arvi{Ar\,{\sc vi}}

\def\caiv{Ca\,{\sc iv}}
\def\cav{Ca\,{\sc v}}
\def\cavi{Ca\,{\sc vi}}
\def\cavii{Ca\,{\sc vii}}
\def\caviii{Ca\,{\sc viii}}

\def\feii{Fe\,{\sc ii}}

\def\nh{\ifmmode n_\mathrm{\scriptstyle H} \else $n_\mathrm{\scriptstyle H}$\fi}
\def\ne{\ifmmode n_\mathrm{\scriptstyle e} \else $n_\mathrm{\scriptstyle e}$\fi}
\def\Qh{\ifmmode Q_\mathrm{\scriptstyle H} \else $Q_\mathrm{\scriptstyle H}$\fi}
\def\Uh{\ifmmode U_\mathrm{\scriptstyle H} \else $U_\mathrm{\scriptstyle H}$\fi}
\def\Nh{\ifmmode N_\mathrm{\scriptstyle H} \else $N_\mathrm{\scriptstyle H}$\fi}
\def\uh{\ifmmode U_\mathrm{\scriptscriptstyle H} \else $U_\mathrm{\scriptscriptstyle H}$\fi}
\def\Uhhp{\ifmmode U_\mathrm{\scriptstyle H,HP} \else $U_\mathrm{\scriptstyle H,HP}$\fi}
\def\Nhhp{\ifmmode N_\mathrm{\scriptstyle H,HP} \else $N_\mathrm{\scriptstyle H,HP}$\fi}
\def\Uhvhp{\ifmmode U_\mathrm{\scriptstyle H,VHP} \else $U_\mathrm{\scriptstyle H,VHP}$\fi}
\def\Nhvhp{\ifmmode N_\mathrm{\scriptstyle H,VHP} \else $N_\mathrm{\scriptstyle H,VHP}$\fi}
\newcommand{\pme}[2]{$^{+#1}_{-#2}$}

\def\Zsun{\ifmmode {\rm Z}_{\odot} \else Z$_{\odot}$\fi}
\def\Msun{\ifmmode {\rm M}_{\odot} \else M$_{\odot}$\fi}
\def\kms{\ifmmode {\rm km~s}^{-1} \else km~s$^{-1}$\fi}
\def\Lya{\ifmmode {\rm Ly}\alpha \else Ly$\alpha$\fi}
\def\Lyb{\ifmmode {\rm Ly}\beta \else Ly$\beta$\fi}
\def\Lyg{\ifmmode {\rm Ly}\gamma \else Ly$\gamma$\fi}
\def\Lyd{\ifmmode {\rm Ly}\delta \else Ly$\delta$\fi}
\def\neaod{\ifmmode n_\mathrm{\scriptscriptstyle AOD} \else $n_\mathrm{\scriptscriptstyle AOD}$\fi}
\def\necrit{\ifmmode n_\mathrm{\scriptstyle cr} \else $n_\mathrm{\scriptstyle cr}$\fi}
\def\ncr{\ifmmode n_\mathrm{\scriptstyle cr} \else $n_\mathrm{\scriptstyle cr}$\fi}
\def\nepi{\ifmmode n_\mathrm{\scriptscriptstyle PI} \else $n_\mathrm{\scriptscriptstyle PI}$\fi}
\def\gtorder{\mathrel{\raise.3ex\hbox{$>$}\mkern-14mu\lower0.6ex\hbox{$\sim$}}}
\def\ltorder{\mathrel{\raise.3ex\hbox{$<$}\mkern-14mu\lower0.6ex\hbox{$\sim$}}}
\newcommand{\vy}[2]{#1_{\scriptscriptstyle #2}}

\shorttitle{ }
\shortauthors{Arav et al.}
\shortauthors{}

\begin{document}


\title{HST/COS observations of quasar outflows in the 500 -- 1050\angstrom\ rest-frame: I \\ The most energetic outflows in the universe and other discoveries}

\author{
Nahum Arav\altaffilmark{1,$\dagger$},
Xinfeng Xu\altaffilmark{1},
Timothy Miller\altaffilmark{1},
Gerard A. Kriss\altaffilmark{2},
Rachel Plesha\altaffilmark{2},
}

\affil{$^1$Department of Physics, Virginia Tech, Blacksburg, VA 24061, USA\\
$^2$Space Telescope Science Institute, 3700 San Martin Drive, Baltimore, MD 21218, USA\\
\hspace{07mm}\
}

\altaffiltext{$\dagger$}{Email: arav@vt.edu}

\begin{abstract}
The Hubble Space Telescope/Cosmic Origins Spectrograph (COS) has opened a new discovery space for studying quasar absorption outflows and their contribution to AGN feedback. Specifically, COS provides high-quality far-ultraviolet (FUV) spectra covering the diagnostic-rich 500--1050 \AA\ rest frame (hereafter, EUV500) of medium redshift objects. The quality and quantity of EUV500 diagnostic troughs allow us to probe the very high-ionization phase, which carries 90\% or more of the outflowing material, as well as to determine the distance of most outflows from the central source ($R$). The first objective is impossible to achieve with ground-based spectra, and $R$ can be measured in only $\sim$1\% of them.
Here, we summarize the main results of the first dedicated survey of such outflows, including the following: \\
1)	Measurements of the three most energetic outflows to date, which can be the main agents for AGN feedback processes in the environments of the host galaxies. \\
2)	All the outflows have a very high-ionization component, similar to the one found in warm absorbers, which carries most of the outflow's kinetic luminosity.  This finding suggests that all the high-ionization outflows observed from the ground also have a similar undetected very high-ionization component. \\
3)	Of the 13 studied EUV500 outflows, 9 have $100<R<2000$ parsecs, 2 have $5<R<20$ parsecs, 1 has $0.05<R<50$ parsecs, and in 1 case, $R$ cannot be determined.\\
4)	One of the outflows has the largest velocity shift (1550 \kms) and acceleration (1.5 cm s$^{-2}$) measured to date. This outflow is physically similar to the fast X-ray outflow detected in quasar PG 1211+143.


\end{abstract}


\section{INTRODUCTION}
\label{sec:Intro}

Quasars show ubiquitous outflows \cite[$\simeq$ 20--50\% of all AGN: e.g.][]{Hewett03,Dai08,Ganguly08,Knigge08}, where blueshifted absorption lines, from ionized material, are attributed to subrelativistic ($\sim10^3-10^4$ \kms) mass ejection. These outflows are prime candidates for producing various AGN feedback processes: curtailing the growth of the host galaxy \cite[e.g.,][]{Ciotti09,Hopkins09,Faucher-Giguere12,Zubovas14,Schaye15,Choi17,Peirani17}, explaining the relationship between the masses of the central black hole and the galaxy's bulge \cite[e.g.][]{Silk98,Blandford04,Hopkins09,Ostriker10,Dubois14,Rosas-Guevara15,Volonteri16,Angles-Alcazar17,Yuan18}, and intercluster medium (ICM) and intergalactic medium (IGM) chemical enrichment \cite[e.g.,][]{Scannapieco04,Khalatyan08,Tornatore10,Barai11,Taylor15,Thompson15}.  Theoretical models indicate that the kinetic luminosity ($\dot{E}_k$) must exceed either 0.5\% \cite[][]{Hopkins10} or 5\% \cite[][]{Scannapieco04} of the quasar's Eddington luminosity ($L_{\mathrm{Edd}}$) for strong AGN feedback to occur; for a more detailed treatment, see \citet{Harrison18}. 

In this paper, we concentrate on the common outflows seen in the rest-frame ultraviolet (UV) portion of the quasar spectra, whose troughs arise from ionized material. We note that AGN outflows are also detected in different phases \citep[e.g., molecular; see][]{Cicone18}, as well as in optical emission lines \citep[e.g., ][]{Zakamska14}, and X-ray \citep[e.g., ][]{Behar17}. Hereafter, we use the term ``quasar outflows" in the narrow sense of rest-frame UV absorption outflows. 

A few of the more important empirical questions regarding quasar outflows are:
\begin{enumerate}
	\item What is their origin and acceleration mechanism?
	\item What is the connection between the outflow and other parts of the AGN phenomenon: accretion disk, broad emission line region, narrow emission line region?
	\item Are the outflows the main agent for the quasar mode of AGN feedback?
\end{enumerate}
To advance our understanding on these questions, it is necessary to determine the outflows' distances from the central source  ($R$), their mass outflow rate ($\dot{M}$) and $\dot{E_{k}}$. 

\begin{figure}[h]

\centering
	\includegraphics[angle=0,scale=0.3]{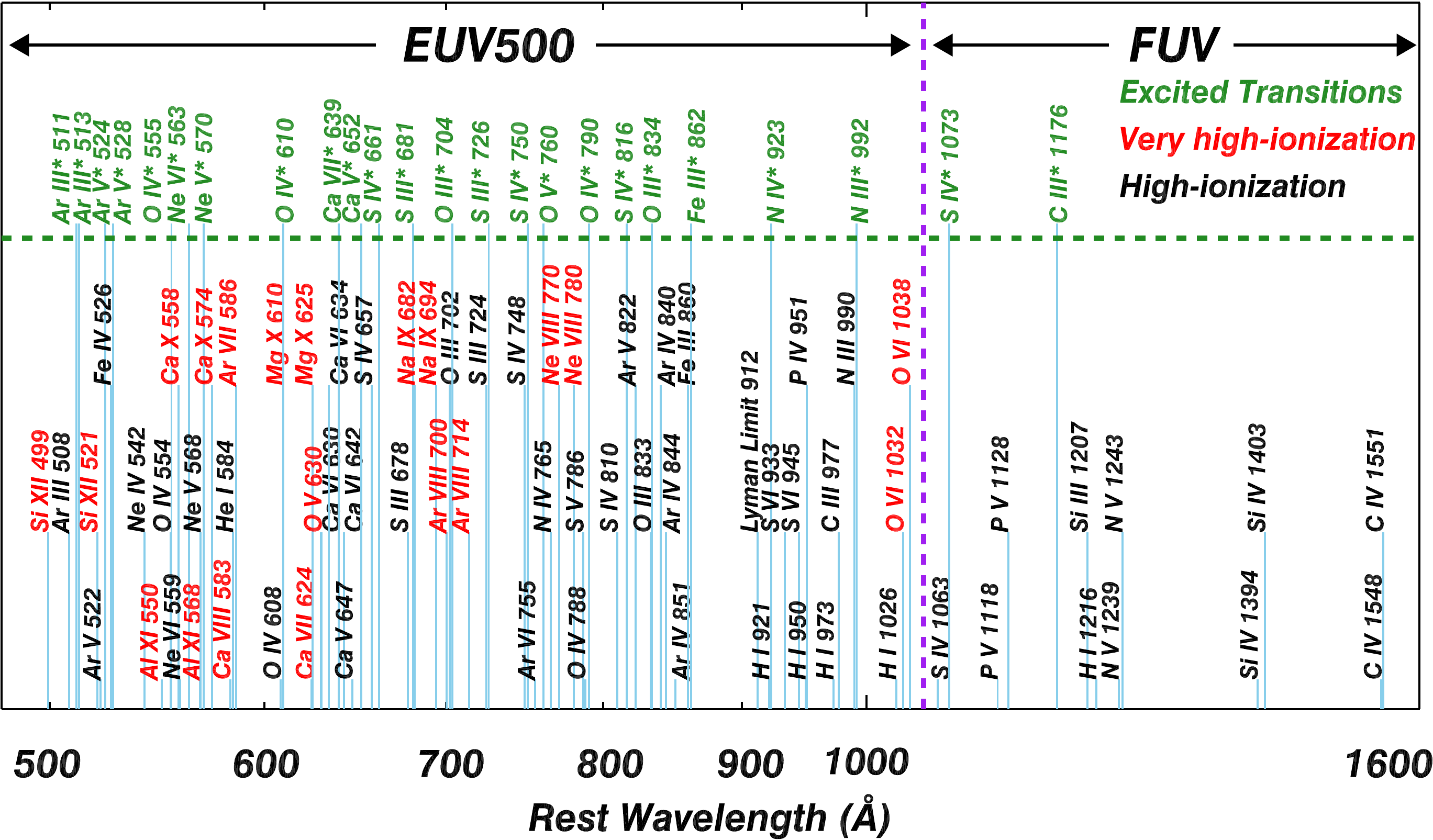}

\caption{Comparison between the diagnostic power of the EUV500 and the $\lambda_{\rm rest}>1050$~\AA\ (FUV) spectral regions.  We show ionic transitions that are observed as absorption troughs in quasar outflows. No troughs from very high-ionization species exist in the FUV, and therefore, 90\% or more of the outflow's column density is unobservable.  In contrast, the EUV500 contains detected outflow troughs from VHI species associated with the 19 transitions shown in red, which arise from 13 different ions (and this is not a complete list). In green, we show transitions from excited states. For high-ionization outflows, only two multiplets produce detected troughs from excited states at $\lambda_{\rm rest}>1050$~\AA. Such detections are quite rare and are difficult to analyze (see section \ref{subsec:advantages}). In contrast, 
for the EUV500, we show a representative sample of 22 transitions from excited states that are expected to produce outflow troughs. We detected such troughs from 16 of these transitions.  
}
\label{fig:EUV500lines}
\end{figure}

The large majority of quasar outflows show absorption troughs from only high-ionization species (e.g., \civ\ and \siiv), and there are more than 10,000 ground-based spectra of such outflows. However, almost all of these only cover rest-frame wavelengths longer than 1050~\AA, where it is very rare to find diagnostic troughs that allow us to measure the distance of the outflow from the central source ($R$) and none for measuring its total hydrogen column density (\Nh). 
The 500--1050~\AA\  rest-frame region (hereafter EUV500) contains an order of magnitude more diagnostic troughs  (see figure 1). These include troughs from very high-ionization (VHI) species (ions with an ionization potential above 100 eV, e.g., \neviii, \naix, \mgx\ and \sixii) whose ionization phase carries most of the outflowing \Nh, \cite[e.g.,][]{Arav13} and troughs  that  allow us to determine $R$, which when combined with \Nh, yields $\dot{E}_k$.  We elaborate on the comparison between diagnostic power of the EUV500 and the $\lambda_{\rm rest}>1050$~\AA\ spectral-regions in section \ref{subsec:advantages} (see also fig.~\ref{fig:EUV500lines}).

From the ground, the EUV500 is accessible for quasars at redshift $z\gtrsim3-4$. Unfortunately, the \Lya\ forest at these redshifts is too thick to allow a meaningful outflow analysis. 
From space, the Hubble Space Telescope (HST) can observe the EUV500 in quasars at the 0.5 $< z <$ 1.5 range. The Cosmic Origins Spectrograph \cite[COS,][]{Green12} on-board HST yields data with sufficient signal-to-noise ratio (S/N) and spectral-resolution to extract outflow science from these quasars.
In \cite{Arav13}, we analyzed COS EUV500 observations of the outflows seen in quasar 
HE~0238-1904 and found that more than 90\% of the outflows' column density --- and therefore $\dot{E}_k$ --- resided in the VHI phase of these outflows.

In order to realize the science potential of EUV500 data for quasar outflows, we executed the first dedicated survey. This spectroscopic survey program (HST GO-14777, PI: Arav) observed 10 quasars with known EUV500 outflows at redshift $z \sim1$.  Each object was observed for four orbits, two each with COS gratings G130M and G160M (the objects and their observations are listed in Table~\ref{tb:observations}). From the 10 targeted objects, we published results for the lowest redshift one \cite[where the highest ionization species detected is \ovi,][]{Miller18}. 

Here (Paper I), we give a  summary of our scientific results and discoveries for the four objects where the outflow data lend themselves to accurate physical analysis. We then discuss the importance of these findings to various
aspects of quasar outflow research.
Detailed analyses of these observed outflows are given in the following series of papers: \\
Paper II \cite[][]{xu20a} gives the full
analysis for 4 outflows detected in SDSS J1042+1646, including the
largest $\dot{E}_k$ ($10^{47}$~erg~s$^{-1}$) outflow measured to date
at $R=800$~pc, and an outflow at $R=15$~pc. \\
Paper III \cite[][]{mil20a} analyzes 4 outflows
detected in 2MASS J1051+1247, which show remarkable similarities, are
situated at $R\sim400$~pc, and have a combined $\dot{E}_k=10^{46}$ erg
s$^{-1}$. \\
Paper IV \cite[][]{xu20b} presents the largest
velocity shift and acceleration measured to date in a broad absorption line (BAL) outflow.  \\
Paper V \cite[][]{mil20b} analyzes 2 outflows
detected in PKS J0352-0711: one outflow at $R=500$~pc and a second
outflow at $R=10$~pc that shows an ionization-potential-dependent
velocity shift for troughs from different ions. \\
Paper VI \cite[][]{xu20c} analyzes 2 outflows
detected in SDSS J0755+2306, including one at $R=1600$~pc with
$\dot{E}_k=10^{46}$~erg~s$^{-1}$. \\

In the remaining five objects, we found significant intervening absorption, including cases of partial Lyman limit systems, as well as cases of significant self-blending of outflow troughs. Since these are somewhat more challenging to analyze, we will present their results in Paper VII (Miller et al.\ 2020c, in preparation).\\

  \begin{deluxetable*}{lllllllllr}
	\scriptsize
	\centering
	\hspace{-1cm}
\tablecaption{Objects observed in program \textit{HST} GO-14777}
\tablehead{
Object & Redshift & \multicolumn{2}{c}{G130M Observation} & \multicolumn{2}{c}{G160M Observation} & Publication & \multicolumn{3}{c}{Identification Observation}\\
& & \multicolumn{1}{c}{Date} & Exp$^a$ & \multicolumn{1}{c}{Date} & Exp$^a$ & & Instrument & Date & Exp$^a$
}
\startdata
LBQS J1206+1052 & 0.396 & 2017 Jul 18 & 4320 & \ \ \   2017 Jul 18 & 4640 & Miller$^b$  & COS G130M & 2010 May 8 & 4840\\
VV2006 J0755+2306 & 0.854 & 2017 Sep 18-19 & 3550 & \ \ \ 2017 Sep 18 & 4660 & Paper VI & COS G140L & 2010 Dec 20 & 900\\
2MASS J1436+0727 & 0.894 & 2017 Jul 17 & 4130 & \ \ \ 2017 Jul 17 & 4660 & Paper VII & COS G140L & 2011 May 11 & 900\\
VV2006 J1329+5405 & 0.950 & 2017 Sep 30 & 3690 & \ \ \ 2017 Sep 30 & 4660 & Paper VII & COS G140L & 2011 Feb 21 & 900\\
PKS J0352--0711 & 0.966 & 2017 Aug 5 & 4070 & \ \ \ 2017 Aug 5 & 4660 & Paper V & FOS G270H & 1993 Sep 26 & 560\\
SDSS J1042+1646 & 0.978 & 2017 Nov 13 & 3360 & \ \ \ 2017 Nov 13 & 4920 & Papers II/IV & COS G140L & 2011 Jun 15 & 900\\
7C J1631+3930 & 1.025 & 2017 May 13 & 4200 & \ \ \ 2017 May 14 & 5200 & Paper VII & FOS G270H & 1993 Jul 21 & 400\\
SDSS J0936+2005 & 1.183 & 2017 Nov 21 & 4360 & \ \ \ 2017 Nov 22 & 4660 & Paper VII & COS G140L & 2011 Feb 13 & 900\\
SDSS J1051+1247 & 1.283 & 2018 Jan 4 & 3460 & \ \ \ 2018 Jan 4 & 4640 & Paper III & COS G130M & 2013 May 17 & 10870\\
SDSS J1123+0137$^c$  & 1.472 & 2017 Nov 15 & 3870 & \ \ \ 2017 Nov 15-16 & 4660 & Paper VII & COS G140L$^{d}$ & 2014 Jun 19 & 4990\\
\hline
\multicolumn{2}{l}{$^a$ Exposure time in seconds}\\
\multicolumn{3}{l}{$^b$ Miller et al. (2018)}\\
\multicolumn{3}{l}{$^c$ Also known as UM 425}\\
\multicolumn{8}{l}{$^d$ Additional identification observations from:  HST FOS G270H: 1994 Nov 8, 1320 sec and SDSS: 2002 Dec 28, 2700 sec}
\enddata
\label{tb:observations}
\end{deluxetable*}

This paper is organized as follows. Section \ref{sec:SAMPLEandID} describes the observed sample.
Section~\ref{sec:path} gives an overview of how the scientific results are extracted from the data. Section~\ref{sec:results} gives a summary of the results from these investigations. In section~\ref{sec:discussion}
we discuss the importance of these findings to various
aspects of quasar outflow research, including: \\
The many advantages of studying quasar outflows using EUV500 data (section \ref{subsec:advantages}); measuring the dominant very high-ionization phase (VHP) of the outflow (section \ref{subsubsec:VHP-measure}); determining the total \Nh\ and ionization structure of the outflows (section \ref{subsubsec:IE}); outflow distance determinations (section \ref{subsubsec:distance-determination}); abundance determinations (section \ref{subsubsec:Abundances}); comparison with X-ray observations of Seyfert and quasar outflows (section \ref{subsec:xray}); comparison with earlier EUV500 observations of quasar outflows (section \ref{subsec:earlier-obs}) and the BAL definition for the EUV500 (section \ref{sec:BAL}).
We summarize the main findings in section \ref{sec:SUMMARY}.

For our analysis, we adopt a cosmology with $h = 0.696$, $\Omega_m = 0.286$, and $\Omega_\Lambda = 0.714$ and use Ned Wright's Javascript Cosmology Calculator website \cite[][]{Wright06}.

\section{THE OBSERVED SAMPLE AND OUTFLOW IDENTIFICATION}
\label{sec:SAMPLEandID}
\subsection{The Observed Sample}\label{subsec:sample}
Details about our targets, including the dates and exposure times for each quasar are found in table \ref{tb:observations}.
Our objects were selected following an exhaustive search in the UV quasar  archives of HST (https://archive.stsci.edu/hst/), where we looked at data from all three generations of UV spectrographs: faint object spectrograph (FOS), space telescope imaging spectrograph (STIS) and cosmic origins spectrograph (COS). Most of these  quasar spectra come from programs whose
primary intent was to probe the intervening absorption from the IGM, the circumgalactic medium (CGM), galaxy halos, or high-velocity clouds. For most of these programs, quasar outflow troughs are a contaminant, and, therefore, objects with known outflows were purposefully omitted from their samples. This explains the paucity of outflow targets in the HST archive.

The sample selection was as follows: 
\begin{enumerate}
\item  The redshift range was 0.5 -- 1.5. At $z > 0.5$ the strong lines from \oiv, \niv\ and \neviii\ are in the observed band of HST; and at $z > 1.5$, the \Lya\ forest becomes thick enough that blending with outflow troughs in these low S/N data becomes a significant concern. 
\item A minimum  continuum flux of $2\times10^{-15}$ erg cm$^{-2}$ s$^{-1}$ \AA$^{-1}$ to allow for a reliable analysis.
\item Trough selection for the objects observed with either COS G140L or G130M. We searched for at least two troughs at the same velocity. To minimize false positive detections of intervening systems, we required the troughs to be wider than 500~\kms\ at a residual intensity of $I=0.9$. For spectra that covered 700~\AA~$<\lambda_{\rm rest}< 800$ \AA, we searched for the strongest expected pairs of resonance lines from the high-ionization phase (HP), \oiv~787~\AA\ and \niv~765~\AA, and/or the \neviii~$\lambda\lambda$770,780 for the VHP. For higher redshift objects, we used the \mgx\ doublet for the VHP and \oiv~608~\AA\
and \ov~630~\AA\ for the HP. Eight targets
from our survey were identified using this procedure. 
\item Trough selection for the objects observed with FOS G270H: For objects
at $0.5<z<1.5$, FOS G270H observations cover only some of the 
880~\AA $<\lambda_{\rm rest}<2180$~\AA\ range. 
In these cases, we searched for at least two troughs at the same velocity from the ``traditional" outflow trough transitions: \ovi$~\lambda\lambda$1031.93,1037.62, 
 \Lya, \nv$~\lambda\lambda$1238.82,1242.80, \siiv$~\lambda\lambda$1393.76,1402.77, and \civ$~\lambda\lambda$1548.20,1550.77.  After identifying candidates in this way, we checked that their \textit{Galaxy Evolution Explorer} (GALEX) photometry in the far-ultraviolet (FUV) is equal to or above the equivalent  flux stated in criterion 2 above. Two targets
from our survey were identified using this procedure.
\end{enumerate}
Selection criteria 3 and 4 prevented bias towards\\
a) Either the HP or the VHPs as we chose objects that showed troughs from either phase. \\
b) Any particular $R$ scale since we searched for only resonance lines.

\subsection{Outflow Identification}\label{subsec:OUTFLOWID}
We define an outflow by the following two criteria. 1) An absorption feature that is at least 500 \kms\ in width at a residual intensity of $I=0.9$. This is similar to the definition of a mini-BAL  \citep{Hamann04}.
Such a width will avoid the vast majority of absorption troughs due to the Milky Way interstellar medium (ISM) and any other intervening absorption systems (e.g., IGM) as these are considerably narrower. 2)  We require at least two troughs from different
transitions that are at the same velocity for an outflow identification.

\begin{figure}[h]
\centering
	\includegraphics[angle=0,scale=0.34]{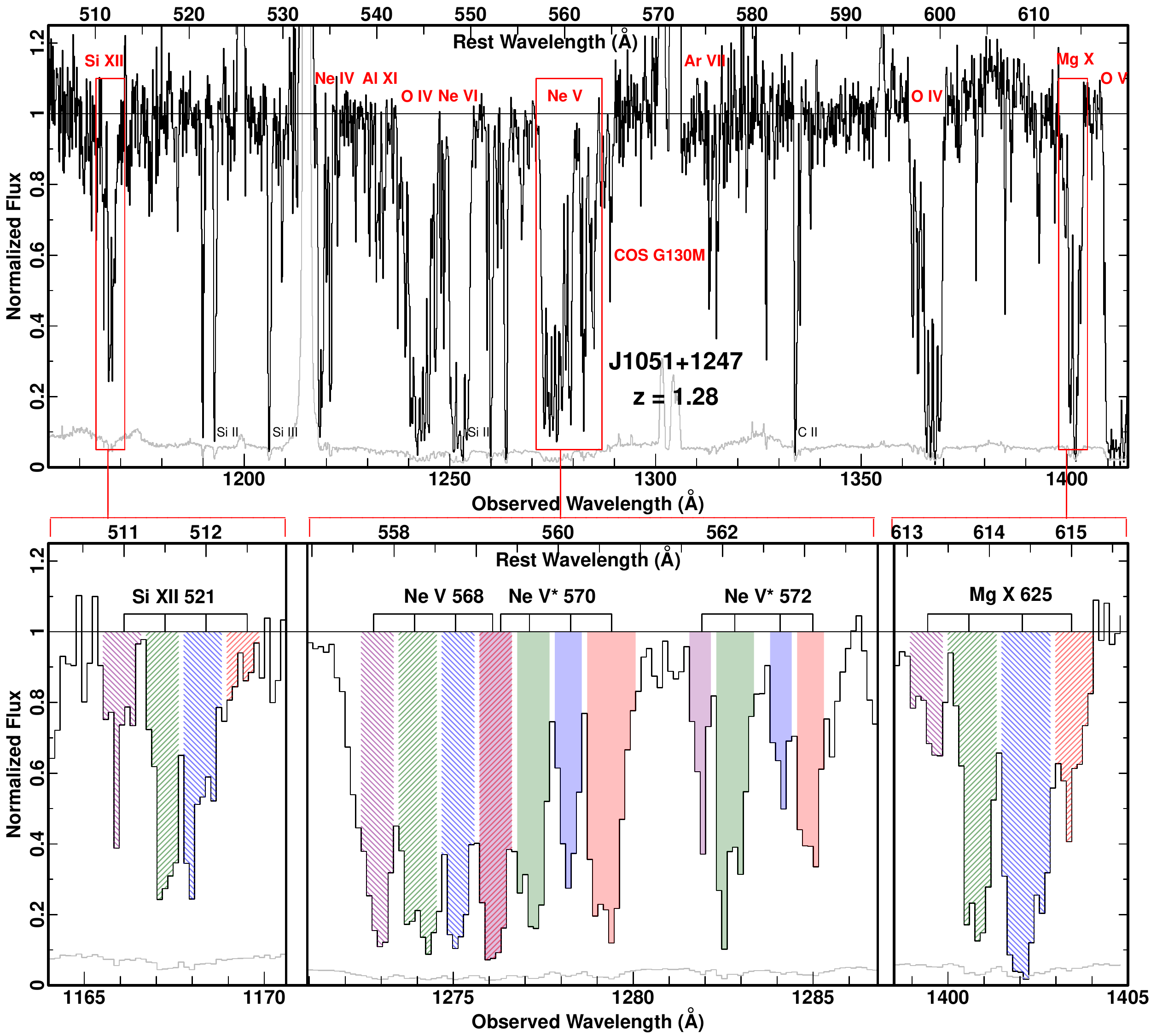}
\caption{\textbf{Top:} a portion of the COS G130M  data for SDSS~J1051+1247, showing troughs from: VHI species (\mgx, \alxi and \sixii), excited transitions (\oiv* and \nev*) and high-ionization species (\oiv). \textbf{Bottom:} a blow up of the \sixii, \nev, and \mgx\ troughs, showing the four detected outflow systems, resonance troughs as hatched fill regions, and troughs from excited states as solid filled regions.}
\label{fig:1051spectrum}
\end{figure}

We define an outflow system as either: a) an outflow where the red side of at least one trough returns to $I=1$, and where the blue side of at least one trough (not necessarily the same trough) returns to $I=1$, or b) an internal structure within a trough where the  maximum separating $I$ is at least 20\% higher than the minima on both sides in at least one trough. An example of one outflow separated into four systems is shown in figure \ref{fig:1051spectrum}. We label systems in ascending order of absolute velocity S1, S2, etc.

\section{OVERVIEW OF SCIENCE EXTRACTION}
\label{sec:path}
We reduced the  data and estimated the errors following the same procedure described in \citet{Miller18}. Detailed analyses of the EUV500 observations from this program are given in Papers II--VII.
Here, we give an overview of the process needed to extract the scientific results from these data. To this end, we use one outflow (S2 in SDSS~J1051+1247, see Paper III) as an example.

\subsection{Richness of Diagnostic Troughs}\label{subsec:Richness}

A portion of the data for SDSS~J1051+1247 (COS G130M observations) is shown in the top panel of figure \ref{fig:1051spectrum}, where we detect 
 troughs from 10 transitions. These transitions include VHI species of \mgx, \alxi\ and \sixii; excited state transitions from \nev\ and \nevi, which are used to determine $R$; and troughs similar to the majority detected at $\lambda_{\rm rest}>1050$~\AA\ (i.e., from resonance transitions of high-ionization species, here \oiv). 
In the full data set, which includes our COS G160M observations of this object, we detect troughs from 17 transitions.

Ionic column densities ($N_{ion}$) are extracted using standard techniques (see section 3.1 in Paper III). On the bottom panel of figure \ref{fig:1051spectrum}, we show a blow up of the \sixii, \nev, and \mgx\ troughs, which are marked in red rectangles on the top panel, for the four detected outflow systems (S2 is marked with blue shading). The keys for determining the physical properties of the VHP of the outflow are the $N_{ion}$ extracted from the \sixii\ and \mgx\ troughs, the \alxi\ trough (seen in the top panel of figure \ref{fig:1051spectrum}), and the \naix\ and \neviii\ troughs detected in the COS G160M data.
Likewise, the keys for determining the electron number density (\ne) of the outflow (and therefore $R$) are the $N_{ion}$ extracted from the troughs associated with the resonance transition \nev~568~\AA\ and the excited state transitions \nev*~570~\AA\ and \nev*~572~\AA\ (see section \ref{subsec:neR} for elaboration).

\subsection{Determining \Nh\ and \Uh}\label{subsec:UHNH}

\begin{figure}[h]
\centering
	\includegraphics[angle=0,scale=0.3]{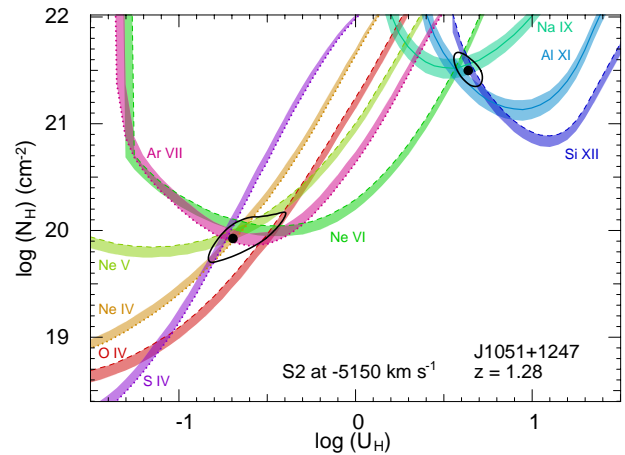}
\caption{Photoionization grid models \cite[based on Cloudy;][]{Ferland17}   showing the outflow's ionization constraints. A solid
		contour represents the locus of \Nh\ and \uh\ models  
		that predict the observed $N_{ion}$. The bands on each side  represent the	1$\sigma$ uncertainties in the measurements. Dashed lines are for $N_{ion}$ lower  limits,  allowing the phase space above them, and similarly dotted lines are $N_{ion}$ upper limits, allowing the phase space below them.
Here, we show the ionization solutions for the $-5150~\kms$ outflow system (S2) seen in the COS data of SDSS~J1051+1247 (portions of which are shown in fig.~\ref{fig:1051spectrum}).  The two solutions for each phase are marked by the black dots and are surrounded by 1$\sigma$ confidence intervals (black contours).		
		}
\label{fig:1051UHNH}
\end{figure}

 Ionization equilibrium in quasar outflows is dominated by photoionization, where the outflow is characterized by its ionization parameter (\Uh) and total hydrogen column density (\Nh).
We run the spectral synthesis code Cloudy \citep[version
c17.00;][]{Ferland17} to generate grids of photoionization simulations \citep[see][]{Arav13} to find the solution that best fits the measured  $N_{ion}$.
The multitude of detected troughs in the EUV500 give many $N_{ion}$ constraints that yield reliable and over-constrained solutions (see section \ref{subsubsec:IE}). Figure \ref{fig:1051UHNH} shows constraints only for S2, whose troughs are shown as the blue filled regions on the bottom panel of figure \ref{fig:1051spectrum}. The data requires two (well-constrained) ionization phases \cite[see][]{Arav13}. We note that the \Nh\ of the VHP
is $\sim$40 times larger than that of the high-ionization phase (HP), and they differ by a factor of 20 in their \uh\ values. Similar results are obtained for the other three outflow systems seen in SDSS~J1051+1247 (see Paper III).

\subsection{Determining \ne\ and $R$}\label{subsec:neR}

The most robust way for determining $R$ for  quasar absorption outflows is the use of troughs from ionic excited states \cite[see section 7.1 in][]{Arav18}. The column density ratio between the excited and resonance states yields the electron number density (\ne). With the knowledge of \ne, we determine $R$ from the value and definition of the ionization parameter \citep[see equation (13.6) in][]{Osterbrock06}: 
\begin{equation}
\label{eq:R}
U_H\equiv \frac{Q_H}{4\pi R^2\vy{n}{H} c}
\end{equation}
where $\vy{n}{H}$~is the hydrogen number density with $\vy{n}{H} \simeq 0.8\vy{n}{e}$ for a highly ionized plasma, $R$ is the distance of the outflow from the central source, $c$ is the speed of light, and $Q_H$ is the incident ionizing photon rate of hydrogen. A comparative discussion of all the methods (found in the literature) that are used to determine $R$ is given 
in \cite{Arav18}'s section 7.1.

The EUV500 includes several transitions from excited states of abundant high-ionization species (e.g., \oiii, {\sc iv}, and {\sc v}; and \nev\ and {\sc vi}, see figure~\ref{fig:EUV500lines}), which produce troughs in most high-ionization outflows.
Thus, \ne\ can be measured in the majority of EUV500 outflows and, in some instances, by more than one diagnostic, which makes the $n_e$ determination more reliable.

\begin{figure}[htp]
\centering
	\includegraphics[angle=0,scale=0.33]{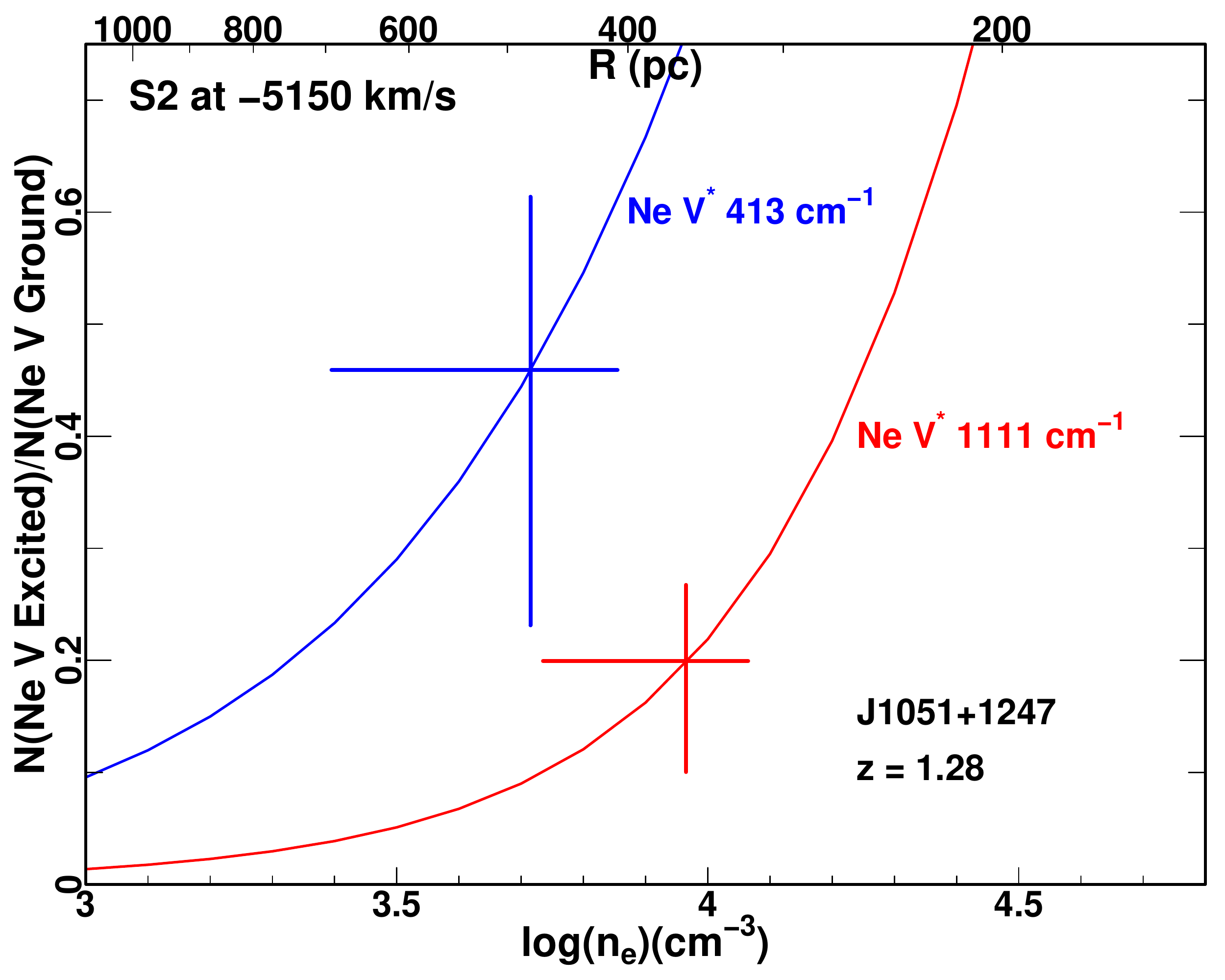}
\caption{ \ne\ and $R$ diagnostics for S2 in SDSS~J1051+1247:  We first determine $n_e$ by measuring the column density ratio between troughs from excited and ground state energy levels of a given ion. The theoretical curves for two \nev\ excited states and the measurement of their $N_{\rm ion}$ ratios with respect to the resonance state are plotted. The top x-axis shows the derived $R$ using equation (\ref{eq:R}) and the \Uh\ value of system 2. }
\label{fig:1051neR}
\end{figure}

The \ne\ measurements for S2 are based on the \nev\ troughs seen on the bottom panel of figure \ref{fig:1051spectrum}. The solid filled troughs are from the excited state transitions at 572.33~\AA\ (E$_{low}$~=~1111~cm$^{-1}$) and 569.83~\AA\ (E$_{low}$~=~413~cm$^{-1}$), where E$_{low}$ is the lower energy level at which the electron absorbs the photon. The hatched filled troughs are from the resonance line at 568.41~\AA\ (E$_{low}$~=~0).
We use the code CHIANTI \citep[version 7.1.3;][]{Dere97,Landi13} to calculate the theoretical $N_{\rm ion}$ ratios of the excited to the E$_{low}$~=~0 states, as a function of \ne. The input temperature is the one determined by Cloudy for the lower ionization solution shown in  figure \ref{fig:1051UHNH}, which produces almost all the \nev\ ions.
Figure \ref{fig:1051neR} shows these theoretical $N_{\rm ion}$ ratios, where the measured ratios with uncertainties for each level are overlaid on the theoretical curves, thus determining \ne.

 For SDSS~J1051+1247, $Q_H=7.3\times 10^{56}$~s$^{-1}$ (see section 4.1 in Paper III).
Using the \ne\ derived from figure \ref{fig:1051neR} and the $U_H$ deduced from the HP solution (the VHP produces negligible amounts of \nev),
we derive $R=360^{+130}_{-100}$ pc for S2 (see elaboration in Paper III).

\subsection{Determining $\dot{M}$ and $\dot{E}_k$ of the Outflow}\label{subsec:Ek}

With the knowledge of \Nh\ and $R$, the  mass outflow rate ($\dot{M}$) and  kinetic luminosity ($\dot{E}_k$) of the outflow can be determined from 
\cite[for elaboration, see section 2.1 in][]{Arav13}:

\begin{equation}
\dot{M}\simeq 4\pi \Omega R \vy{N}{H}\mu m_p v 
\label{eqn:mflux}
\end{equation}
and 
\begin{equation}
  \dot{E}_k=\frac{1}{2}\dot{M}v^2\simeq 2\pi \Omega R \vy{N}{H}\mu m_p v^3
\label{eqn:Edotk}
\end{equation}
where $\Omega$ is the fraction of the total solid angle occupied by the outflow,  $m_p$ is the mass of the proton, $\mu\simeq1.4$ is the mean molecular weight of the plasma per proton, and $v$ is the velocity of the outflow. $\Omega=0.40$ is estimated by the fraction of all quasars showing similar width outflows (see section 4.1 in Paper III), and $v$ is measured from the deepest portion of the outflow trough. We note that since in all our outflows $\Delta v/v<0.1$ (where  $\Delta v$ is the width of the outflow), the exact position where $v$ is measured in the outflow produces only small differences in the derived $\dot{M}$
and $\dot{E}_k$.

Using the total measured $\vy{N}{H}$  and the above $R$ value,  
we obtain for S2 $\dot{E}_k\simeq2.9\times10^{45}$ erg s$^{-1}$. The four outflow systems in 
SDSS~J1051+1247 combined have $\dot{E}_k\simeq 8.8\times10^{45}$ erg s$^{-1}$, which is roughly 7\% of its $L_{\mathrm{Edd}}$.  Therefore, 
the  outflows in this object can be major contributors to AGN
feedback mechanisms.

\subsection{Volume Filling Factor}
\label{sec:2Phases:Factor}
Kinematic similarities (both velocity centroid and width) between troughs from the HP and VHP are evidence that the two phases are cospatial. 
The volume filling factor, $\vy{f}{V}$, between the two phases is defined as the ratio of volumes between the HP and VHP \cite[see section 8.1 in][]{Arav13}. For each phase, the volume is proportional to \Nh/$\vy{n}{H}$, and the $\vy{n}{H}$ ratio between the HP and the VHP is given by $U_{\text{H},\scriptscriptstyle{VHP}}/U_{\text{H},\scriptscriptstyle{HP}}$. Therefore,

\begin{equation}\label{eq:Vfactor}
\begin{split}
\vy{f}{V} \equiv\ \frac{V_{\scriptscriptstyle{HP}}}{V_{\scriptscriptstyle{VHP}}} = \frac{N_{\text{H},\scriptscriptstyle{HP}}}{N_{\text{H},\scriptscriptstyle{VHP}}} \times \frac{U_{\text{H},\scriptscriptstyle{HP}}}{U_{\text{H},\scriptscriptstyle{VHP}}}
\end{split}
\end{equation}
For S2, we obtain  $\vy{f}{V}=0.001$.

\section{RESULTS}
\label{sec:results}

Table \ref{tb:ParaSystems} shows all the derived parameters for the 13 outflows detected in the EUV500 observations of the four quasars discussed here.
For the photoionization solutions, we systematically used the HE0238 SED \cite[][]{Arav13}. For SDSS J1042+1646 and SDSS J1051+1247, we used solar abundances.
For PKS J0352--0711 and  VV2006 J0755+2306, the photoionization solutions require roughly 5 times solar metallicity (see Papers V and VI). For comparison, we also show results for the previous largest $\dot{E_{k}}$ outflows \cite[][]{Arav13,Chamberlain15b}. 
In this section, we elaborate about the major findings and their significance to quasar outflow research.

\begin{turnpage}
	\begin{deluxetable*}{c c c c c c c c c c c  l}[htb!]
		\tablewidth{1.0\textwidth}
		\tabletypesize{\small}
		\tablecaption{Analysis Results$^{(a)}$}
		\tablehead{
			\colhead{System}   & \colhead{ $v$$^{(b)}$} & \colhead{log(\Uhhp)} & \colhead{log(\Nhhp)} & \colhead{log(\ne)}& \colhead{log(\Uhvhp)} & \colhead{log(\Nhvhp)} & \colhead{log($\vy{f}{V}$)$^{(c)}$}& \colhead{R} & \colhead{$\dot{M}$}& \colhead{Log $\dot{E_{k}}$} & \colhead{ $\dot{E_{k}}/L_{Edd}$}
			\\ [-2mm]
			\\
			\colhead{}   & \colhead{(km s$^{-1}$)} & \colhead{log}& \colhead{log(cm$^{-2}$)} & \colhead{log(cm$^{-3}$)}& \colhead{log}  & \colhead{log(cm$^{-2}$)} &\colhead{}   & \colhead{pc} & \colhead{($M_{\odot}$ yr$^{-1}$)} & \colhead{log(erg s$^{-1}$)} & \colhead{$\%$}
		}
		
		\startdata
		
		\multicolumn{11}{l}{\textbf{SDSS J1042+1646, L$_{\text{bol}}$ = 1.5 $\times$ 10$^{47}$ erg s$^{-1}$}} \\
		\hline
		\textbf{1a} &\textbf{-4950} &\textbf{-1.0$^{+0.2}_{-0.3}$}&\textbf{20.4$^{+0.4}_{-0.6}$}&\textbf{3.7$^{+0.2}_{-0.3}$} &\textbf{0.4$^{+0.2}_{-0.1}$}&\textbf{22.4$^{+0.2}_{-0.1}$} & \textbf{-3.4$^{+0.5}_{-0.7}$}
		&\textbf{840$^{+500}_{-300}$}&\textbf{2800$^{+200}_{-800}$} &\textbf{46.4$^{+0.1}_{-0.1}$}&\textbf{10$^{+3}_{-2}$}\\
		
		\textbf{1b} &\textbf{-5750} &\textbf{-0.9$^{+0.2}_{-0.2}$}&\textbf{20.5$^{+0.4}_{-0.3}$}&\textbf{3.8$^{+0.2}_{-0.3}$} &\textbf{0.5$^{+0.2}_{-0.2}$}&\textbf{22.5$^{+0.3}_{-0.2}$} & \textbf{-3.4$^{+0.6}_{-0.5}$}
		&\textbf{800$^{+300}_{-200}$}&\textbf{4300$^{+1200}_{-1500}$} &\textbf{46.7$^{+0.2}_{-0.1}$}&\textbf{20$^{+14}_{-4}$}\\
		
		\textbf{2} &\textbf{-7500} &\textbf{-0.6$^{+0.2}_{-0.1}$}&\textbf{20.8$^{+0.3}_{-0.3}$}&\textbf{5.8$^{+0.5}_{-0.3}$} &\textbf{0.4$^{+0.1}_{-0.1}$}&\textbf{22.4$^{+0.1}_{-0.1}$} & \textbf{-2.6$^{+0.4}_{-0.6}$}
		&\textbf{15$^{+8}_{-8}$}&\textbf{81$^{+20}_{-30}$} &\textbf{45.1$^{+0.1}_{-0.2}$}&\textbf{0.5$^{+0.2}_{-0.2}$}\\
		
		\textbf{3} &\textbf{-9940} &\textbf{-0.7$^{+0.1}_{-0.1}$}&\textbf{20.7$^{+0.2}_{-0.2}$}&\textbf{--} &\textbf{0.1$^{+0.1}_{-0.1}$}&\textbf{21.5$^{+0.1}_{-0.1}$} & \textbf{-1.6$^{+0.6}_{-0.3}$}
		&\textbf{--}&\textbf{--} &\textbf{--}&\textbf{--}\\
		
		\textbf{4} &\textbf{-21050\hspace{1.8mm} }&\textbf{ -- }&\textbf{ -- }&\textbf{4.5 -- 10.5} &\textbf{0.2 -- 0.9}&\textbf{20.8 -- 22.2} & \textbf{--}
		&\textbf{0.05 -- 50}&\textbf{0.07 -- 140} &\textbf{43.0 -- 46.3}&\textbf{0.004 -- 8}\\
		
		\hline
		\\ [0.01mm]
		\multicolumn{11}{l}{\textbf{2MASS J1051+1247, L$_{\text{bol}}$ = 1.3 $\times$ 10$^{47}$ erg s$^{-1}$}}\\
		\hline
		\textbf{1} &\textbf{-4900} &\textbf{-0.8\pme{0.3}{0.1}}&\textbf{20.3\pme{0.4}{0.2}}&\textbf{3.9\pme{0.1}{0.3}} &\textbf{0.3\pme{0.5}{0.1}}&\textbf{21.1\pme{0.4}{0.6}} & \textbf{-1.9$^{+0.8}_{-0.7}$}
		&\textbf{460\pme{200}{130}}&\textbf{180\pme{310}{120}} &\textbf{45.1\pme{0.5}{0.5}}&\textbf{1.1\pme{2.4}{0.8}}\\
		
		\textbf{2} &\textbf{-5150} &\textbf{-0.7\pme{0.3}{0.2}}&\textbf{19.9\pme{0.2}{0.2}}&\textbf{4.0\pme{0.1}{0.2}} &\textbf{0.6\pme{0.1}{0.1}}&\textbf{21.5\pme{0.2}{0.2}} & \textbf{-2.9$^{+0.4}_{-0.4}$}
		&\textbf{360\pme{130}{100}}&\textbf{350\pme{260}{170}} &\textbf{45.5\pme{0.2}{0.3}}&\textbf{2.3\pme{3.1}{1.4}}\\
		
		\textbf{3} &\textbf{-5350} &\textbf{-0.3\pme{0.2}{0.6}}&\textbf{20.6\pme{0.5}{0.5}}&\textbf{4.2\pme{0.1}{0.4}} &\textbf{0.6\pme{0.1}{0.1}}&\textbf{21.5\pme{0.2}{0.3}} & \textbf{-1.8$^{+0.6}_{-0.8}$}
		&\textbf{180\pme{220}{50}}&\textbf{180\pme{320}{90}} &\textbf{45.2\pme{0.5}{0.3}}&\textbf{1.3\pme{3.0}{0.8}}\\
		
		\textbf{4} &\textbf{-5650} &\textbf{-0.8\pme{0.3}{0.1}}&\textbf{19.8\pme{0.2}{0.2}}&\textbf{3.9\pme{0.1}{0.2}} &\textbf{0.6\pme{0.1}{0.1}}&\textbf{21.3\pme{0.3}{0.3}} & \textbf{-2.9$^{+0.5}_{-0.4}$}
		&\textbf{460\pme{160}{140}}&\textbf{300\pme{380}{170}} &\textbf{45.5\pme{0.3}{0.4}}&\textbf{2.3\pme{4.4}{1.5}}\\
		
		\hline
		\\ [0.01mm]
		\multicolumn{11}{l}{\textbf{PKS J0352-0711$^{(d)}$, L$_{\text{bol}}$ = 5.5 $\times$ 10$^{46}$ erg s$^{-1}$}}\\
		\hline
		\textbf{1} &\textbf{-1950} &\textbf{-0.7$^{+0.3}_{-0.3}$}&\textbf{19.1$^{+0.5}_{-0.9}$}&\textbf{3.2$^{+0.2}_{-0.1}$} &\textbf{0.2$^{+0.6}_{-0.1}$}&\textbf{20.3$^{+0.4}_{-0.6}$} & \textbf{-2.1$^{+0.8}_{-1.2}$}
		&\textbf{520$^{+300}_{-150}$}&\textbf{11$^{+22}_{-8}$} &\textbf{43.1$^{+0.5}_{-0.6}$}&\textbf{0.01$^{+0.04}_{-0.01}$}\\
		
		\textbf{2} &\textbf{-3150} &\textbf{-1.0$^{+0.2}_{-0.5}$}&\textbf{20.5$^{+0.4}_{-0.7}$}&\textbf{5.8$^{+0.5}_{-0.3}$} &\textbf{0.3$^{+0.4}_{-0.2}$}&\textbf{21.6$^{+0.3}_{-0.3}$} & \textbf{-2.4$^{+0.6}_{-0.9}$}
		&\textbf{9$^{+5}_{-5}$}&\textbf{7.5$^{+9}_{-5}$} &\textbf{43.4$^{+0.3}_{-0.5}$}&\textbf{0.02$^{+0.06}_{-0.01}$}\\

		\hline
\\ [0.01mm]
\multicolumn{11}{l}{\textbf{SDSS J0755+2306$^{(d)}$, L$_{\text{bol}}$ = 4.4 $\times$ 10$^{46}$ erg s$^{-1}$}}\\
\hline
\textbf{1} &\textbf{-5520} &\textbf{-1.2$^{+0.3}_{-0.2}$}&\textbf{19.9$^{+0.6}_{-0.2}$}&\textbf{4.3$^{+0.2}_{-0.2}$} &\textbf{$>$0.1}&\textbf{$>$20.7} & \textbf{$<$-2.1}
&\textbf{270$^{+100}_{-90}$}&\textbf{$>$21} &\textbf{$>$44.3}&\textbf{$>$0.2}\\

\textbf{2} &\textbf{-9660} &\textbf{-1.6$^{+0.6}_{-0.5}$}&\textbf{19.5$^{+0.7}_{-0.6}$}&\textbf{3.1$^{+0.8}_{-0.5}$} &\textbf{$>$0.1}&\textbf{$>$21.4} & \textbf{$<$-3.3}
&\textbf{1600$^{+2000}_{-1100}$}&\textbf{$>$450} &\textbf{$>$46.1}&\textbf{$>$12.5}\\

\hline
\\ [+0.01mm]
\multicolumn{11}{l}{\textbf{Comparison to Other Energetic Outflows:}}\\
\hline
\textbf{HE 0238--1904$^{(a,e)}$} &\textbf{-5000} &\textbf{-2.4$^{+2.0}_{-0.1}$}&\textbf{17.5$^{+2.0}_{-0.1}$}&\textbf{3.7$^{+0.1}_{-0.1}$} &\textbf{0.5$^{+0.2}_{-0.1}$}&\textbf{20.8$^{+0.1}_{-0.2}$}& \textbf{-6.2$^{+2.8}_{-0.3}$}
&\textbf{3000$^{+900}_{-2800}$}&\textbf{160$^{+80}_{-150}$} &\textbf{45.7$^{+0.2}_{-1.2}$}&\textbf{4$^{+2}_{-3}$}\\

\textbf{J0831+0354$^{(a,f)}$} &\textbf{-10800\;\, }&\textbf{-0.2$^{+0.4}_{-0.5}$}&\textbf{22.4$^{+0.5}_{-0.5}$}&\textbf{4.4$^{+0.3}_{-0.2}$} &\textbf{--}&\textbf{--} & --
&\textbf{80$^{+27}_{-18}$}&\textbf{230$^{+330}_{-130}$} &\textbf{45.9$^{+0.4}_{-0.3}$}&\textbf{8$^{+11}_{-4}$}\\

\vspace{-2.2mm}
\enddata

\tablecomments{\\
(a). The results for the last two entries are from \cite{Arav13} (HE~0238--1904) and \cite{Chamberlain15b} (SDSS~J0831+0354).\\
(b). The velocity centroid of each outflow system. Conservative uncertainties are $\pm40$~\kms\ for all outflows.\\
(c). The volume filling factor of the outflow's high-ionization phase relative to the very high-ionization phase.\\
(d). The photoionization solutions assume roughly five times solar metallicity (see Paper V)\\
(e). For quasar HE 0238--1904, L$_{\text{bol}}$ = 1.5 $\times$ 10$^{47}$ erg s$^{-1}$.\\
(f). For quasar SDSS J0831+0354, L$_{\text{bol}}$ = 6.2 $\times$ 10$^{46}$ erg s$^{-1}$.
}
\label{tb:ParaSystems}
\end{deluxetable*}

\end{turnpage}

\subsection{The Prevalence of the VHP and Its Importance to Outflow Research}
\label{sec:VHP}

\cite{Arav13} showed the existence of two ionization phases in a quasar outflow. One is the HP that is detected in all outflows observed at $\lambda_{\rm rest}>1050$~\AA\ (the large majority of ground-based observations), and the other is the VHP evident by troughs from VHI species (e.g., \neviii, \mgx\ and \sixii) that can be detected only in the EUV500. All 13 outflows we discuss here have a VHP, and our analysis yields the following insights: \\
 
\noindent 1. The VHP carries between 5 and 100 times larger \Nh\ (and, therefore, $\dot{E_{k}}$, see equation (\ref{eqn:Edotk})) than the HP (see Table 2). This demonstrates that a) quantitative studies of the outflows' origin and their effects on the host galaxy's environment are crucially dependent on the parameters of the VHP, and b) attempting to deduce the VHP \Nh\ from the \Nh\ of the HP using a single scale factor would yield inaccurate results since there is a large spread (factor of 20) of the \Nh\ ratio between the phases. \\

\noindent 2.  Twelve of the thirteen outflows have both an HP and a VHP (S4 in SDSS~J1042+1646 shows only a VHP) and have detected \oiv\ troughs.  In a given outflow, the expected optical depth of the \oiv\ troughs at 609~\AA\ ($\tau_{\textrm{\oiv}609}$) and/or 788~\AA\ ($\tau_{\textrm{\oiv}788}$) should be similar to that of the \civ~1549~\AA\ trough ($\tau_{\textrm{\civ}1549}$). This is because, for example,
\begin{equation}
\begin{aligned}
\frac{\tau_{\textrm{\oiv}788}}{\tau_{\textrm{\civ}1549}} =
\left[\frac{N_\textrm{\oiv}}{N_\textrm{\civ}}\right]\left[\frac{f_{\textrm{\oiv}788}\ 788\ \textrm{\AA}}{f_{\textrm{\civ}1549}\ 1549\ \textrm{\AA}}\right] \\= \left[\frac{O}{C}\frac{F_{\textrm{\oiv}}(\textrm{\uh})}{F_{\textrm{\civ}}(\textrm{\uh})}\right]\left[0.29\right] = 0.61\frac{F_{\textrm{\oiv}}(\textrm{\uh})}{F_{\textrm{\civ}}(\textrm{\uh})} \sim 1
\end{aligned}
\end{equation}
where $N_\textrm{\oiv}$ and $N_\textrm{\civ}$ are the ionic column densities for \oiv\ and \civ, respectively; $f$ is the oscillator strength; O/C is the abundance ratio of oxygen to carbon (2.1 for solar metallicity); and $F(\textrm{\uh})$ is the ratio of the ion's number density to the total number density for the element (all ionization stages).
Since $F_{\textrm{\oiv}}(\textrm{\uh})$/$F_{\textrm{\civ}}(\textrm{\uh}) = 1.6$ near the peak $F(\textrm{\uh})$ for both ions, the expected value for 
$\tau_{\textrm{\oiv}788}/\tau_{\textrm{\civ}1549}\simeq 0.61\times1.6 = 1.0$. Therefore, all 12 outflows should have a detectable \civ~1549~\AA\ trough, labeling them as high-ionization outflows if observed only at $\lambda_{\rm rest}>1050$~\AA.

It is also probable that the large majority of high-ionization outflows observed at $\lambda_{\rm rest}>1050$~\AA\ also have a VHP since, a) we detect only 1 of 13 outflows that has only a VHP; and b) \ovi~1034~\AA\ troughs are detected in almost all the ground-based spectra of outflows that cover their expected wavelength range. Since \ovi\ is a VHI (ionization potential of 138 eV), its appearance strongly suggest the existence of a VHP. 

\subsection{The Three Most Energetic Outflows to Date}
\label{sec:energetic-outflows}

Even with errors taken into account, outflows 1a and 1b in SDSS J1042+1646 individually have  an  $\dot{E_{k}}$
larger than the previously published record \cite[][see the last entry in table \ref{tb:ParaSystems}]{Chamberlain15b}.  Outflow 2 in  
SDSS J0755+2306  also has a larger $\dot{E_{k}}$ nominally, but within the errors, the value is similar to the previous record. A similar situation occurs for the combined  $\dot{E_{k}}$ of the four outflows seen in 2MASS J1051+1247, which are probably physically related (see Paper III). 
Therefore, we have three outflows (the combined 1a and 1b in SDSS J1042+1646; the combined 1, 2, 3 and 4 outflow in 2MASS J1051+1247; and outflow 2 in 
SDSS J0755+2306) whose $\dot{E_{k}}$ exceeds or is equal to the largest $\dot{E_{k}}$ value currently reported in the literature.
This plethora of extremely energetic outflows stem from our ability to measure the VHI phase, which carries 80\%-99\% of the total \Nh\ in our studied EUV500 outflows (see table \ref{tb:ParaSystems}).  

We note that the claim of the "Most Energetic Outflows to Date" extends to all quasar outflows where $\dot{E_{k}}$ can be reliably determined (i.e., where $R$ can be measured). This includes all outflow phases at any redshift.

\subsection{Contribution to AGN Feedback}
\label{sec:feedback}

All three outflows described in section \ref{sec:energetic-outflows} have a 
$\Gamma_{Edd}\equiv\dot{E_{k}}/L_{Edd}$  larger than the 5\% conservative threshold needed for an outflow to produce major AGN feedback \cite[][]{Scannapieco04}. 
 Outflows 2 and 4 in SDSS J1042+1646, and outflow 1 in SDSS J0755+2306
have $\Gamma_{Edd}$ values close to 0.5\%, which is the lower threshold for producing significant AGN feedback \cite[][]{Hopkins10}.
The two independent outflows in PKS J0352--0711 have $\Gamma_{Edd}\simeq0.03$\%, which is too small to produce significant AGN feedback.

Therefore, roughly half of the outflows discussed here have a high enough $\Gamma_{Edd}$ value to produce major AGN feedback processes. Although our sample is small and heterogeneous (see section \ref{subsec:sample}),
this finding suggests that a large fraction of quasar outflows are capable of  
producing AGN feedback once the $N_H$ of their very high-ionization phase is taken into account.

\subsection{Distance from the Ionizing Source $(R)$}
\label{sec:}

 Table \ref{tb:ParaSystems} shows that 9 of the 13 studied EUV500 outflows have $100<R<2000$ pc, 2 have $5<R<20$ pc, 1 has $0.05<R<50$ pc and for 1 outflow, $R$ cannot be determined. This spread of $R$ values support the findings from  ground-based observations of \siv\ troughs that about half of all quasar outflows are situated at $R>100$ pc \cite[][]{Arav18,Xu19a}.

\subsection{The Largest Quasar Outflow Acceleration}
\label{sec:}

Outflow 4 in SDSS J1042+1646 (classified as a BAL outflow) has the largest velocity shift (1550 \kms) and average acceleration (1.5 cm s$^{-2}$) measured to date (see Paper IV). These findings are based on two epochs of COS observations, 2011 and 2017, which are separated by 3.2 years in the quasar's rest frame. Between these two epochs, the outflow's velocity changed from --19,500 km s$^{-1}$ to --21,050 km s$^{-1}$. A few points to note about this finding:\\
1. This is the first time where a quasar outflow acceleration is observed in more than one trough (4 troughs show the same velocity shift).\\
2. This is the first time where we have $R$ constraints for an accelerating outflow. Together with future HST/COS observations, these $R$ constraints will allow us to test outflow acceleration models \cite[e.g.,][]{Murray97} in a more substantial way than was possible previously.\\
3. The systematic investigation of \civ\ BAL acceleration/deceleration reported in \cite{Grier16} shows a low detection rate of accelerating outflows (2 out of 140 quasars), where their two acceleration candidates show velocity shifts of up to $\sim$ 900 km s$^{-1}$ over rest-frame times of 3 -- 5 years.
Due to this low detection rate, our finding would have been unusual if observed solely in the \civ\ trough (which is in a wavelength range that our observations do not cover). However, as noted in section \ref{sec:VHP}, this EUV500 outflow is the only one where we detect troughs only from the VHP (and therefore we do not expect to detect a \civ\ trough associated with it). 
The 12 outflows with a HP do not show any velocity shift over two COS epochs that are separated by a similar time interval to those of SDSS J1042+1646 (see Papers II, III, V and VI). The non detection in 12 cases is consistent with the \cite{Grier16} detection rate of 2 in 140 cases.
Therefore, we speculate that perhaps pure VHP outflows show a higher rate of detectable acceleration since they might be situated closer to the central source (in this case $0.05<R<50$ parsecs). This may also explain the high acceleration value for this outflow.

\subsubsection{Similarity with the PG 1211+143 X-Ray Outflow}
\label{sec:UFO}
Outflow S4 has a similar velocity to the X-ray outflow seen in PG 1211+143 (--17,300 km s$^{-1}$), which is detected by troughs from \nex--Ly$\alpha$, \mgxii--Ly$\alpha$, \sixiii--He$\alpha$ and \sixiv--Ly$\alpha$ using \textit{Chandra} observations \citep[][see also \cite{Pounds16a,Pounds16b} for detection of similar troughs using \textit{XMM-Newton} RGS data]{Danehkar18}. This X-ray absorber in PG 1211+143 is well fitted with log(\Nh) $\sim$ 21.5 and log($\xi$) $\sim$ 2.9, where $\xi$ is the X-ray ionization parameter. For the HE 0238 SED, log(\Uh) = log($\xi$) -- 1.3. The UV counterpart of this X-ray outflow has been detected in HST/COS observations, which yields a broad Ly$\alpha$ absorption feature at \textit{v} = --17,000 km s$^{-1}$  \citep[--0.056$c$;][]{Kriss18}.  We compare the $v$, \Nh, and \Uh\ values between the X-ray outflow in PG 1211+143 and outflow S4 in Table 3 of Paper IV. We conclude that our observations in the EUV500 band have probed an outflow with similar physical characteristics to the one observed in PG 1211+143.

\subsection{Ionization-Potential-Dependent Velocity Shift}
\label{sec:shift}
Outflow 2 observed in PKS J0352-0711 (at $v\simeq -$3150~km~s$^{-1}$) shows a unique velocity centroid shift between associated troughs.
Troughs from \niii\ and \oiii\ have a velocity centroid at 
$v\simeq -$3100~km~s$^{-1}$. Troughs from higher ionization potential (IP) species show a gradual velocity-centroid shift, whose magnitude is correlated with increasing IP, where troughs from the highest IP ions (\neviii\ and \naix) have a velocity centroid at $v\simeq -$3200~km~s$^{-1}$.

\section{DISCUSSION}
\label{sec:discussion}

\subsection{The Many Advantages of Studying Quasar Outflows Using EUV500 Data}\label{subsec:advantages}

Here, we detail the necessity of analyzing EUV500 observations and the advantages compared to ground-based data. The only disadvantage of such an analysis is programmatic: the need for HST observations, which is a limited and highly sought after resource in many research fields. 

\subsubsection{Measuring the Dominant VHP of the Outflow}\label{subsubsec:VHP-measure}
As detailed in section \ref{sec:VHP}, the very high-ionization phase (VHP) carries up to 99\% of the column density and energy of the outflow. Clearly, a reliable physical study of the outflows
requires measurements of the VHP phase.  Ground-based observations can detect troughs from the \ovi\ doublet around 1034~\AA\ which is a very high-ionization (VHI) species.  However, measurements of $N_{\rm ion}$ from \ovi\ and lower ionization species do not require a VHP to fit the data. For example, \cite{Xu18} measured  $N_{\rm ion}$(\ovi), yet an ionization solution with only an HP fits the data well.

Therefore, measuring the VHP necessitates EUV500 data covering troughs from \neviii\ and higher ionization species  (e.g., see figure \ref{fig:1051UHNH}).

\subsubsection{Determining the Total \Nh\ and Ionization Structure of the Outflows}\label{subsubsec:IE}
Outflow troughs are known to exhibit non-black saturation, where in extreme cases the real optical depth is a thousand times larger than the apparent optical depth (AOD) deduced from the depth of the trough 
\cite[][]{Borguet12b}. Not accounting for this possibility can lead to \Nh\ values that are only a few percent of the actual one \cite[][]{Arav15}. Therefore, we usually treat the $N_{\rm ion}$
of the abundant species (\hi, \civ, \nv\ and sometimes \siiv) as lower limits. Since these are the only troughs detected in the large majority of ground-based data, it is not possible to derive a \uh, \Nh\ solution for these outflows. 

In contrast, the EUV500 data show many more troughs from ions with a large range of ionic abundances.
From such data we can extract: $N_{\rm ion}$ where we have indications that the trough is not saturated; lower limits for troughs thought to be saturated; and upper limits for non-existent troughs of a given species. We are conservative in our decision to adopt $N_{\rm ion}$ as measurements (see section 3.1 in Paper III). However, the plethora of EUV500 troughs yield enough upper and lower limits to well-constrain both ionization phases of the outflow. 
Moreover, these limits are immune to saturation effects. Lower limits allow for the solution to be above the $N_{\rm ion}$ curve irrespective of the real $N_{\rm ion}$, and upper limits of course do not suffer from saturation.  Given enough constraints of both types, a well-confined solution can be reached.

As an example, in figure \ref{fig:1051UHNH}, the photoionization solutions for the HP in this outflow is constrained only by $N_{\rm ion}$ upper and lower limits.  However, the multitude of upper and lower limits tightly constrain the solution. 

Therefore, it is relatively straightforward to determine the photoionization solutions for EUV500 outflows, whereas it is very challenging for the large majority of outflows with only ground-based spectra.

\subsubsection{Outflow Distance Determinations}\label{subsubsec:distance-determination}

Most published $R$ determinations \cite[e.g.,][]{deKool01,deKool02c,Korista08,Moe09,Bautista10,Dunn10,Aoki11,Hamann01,Leighly18,Lucy14} come
from singly ionized species that have excited levels (mostly from \Siii\ and \feii). However,
some 90\% of quasar outflow spectra show absorption troughs only from  higher
ionization species. Therefore, the applicability of $R$ derived from
\Siii\ and \feii\ to the majority of outflows is somewhat
model-dependent \cite[see discussion in section 1 of][]{Dunn12}.

From the ground, the
main high-ionization species with detected troughs arising
from an excited state is \siv, which has resonance and excited
level transitions at 1063~\AA\ and 1072~\AA, respectively (\ciii* is another option, but there are only a handful of spectra where it is detected).
Published $R$ determinations using \siv\ include \cite{Borguet12a,Borguet13,Chamberlain15b}; and \cite{Xu18,Xu19a}.

However, a few factors limit the use of \siv\ as an $R$ diagnostic:
1) Due to the small oscillator strength of the transitions and the low abundance of S compared to C, these troughs are observed in only 10-15\% of the spectra that show \civ\ troughs \cite[][]{Arav18}.
2) The high redshift needed to shift the \siv\ trough into the ground-based spectral region does not allow the detection of these troughs in quasars at redshift smaller than 2--2.8 (depending on the spectrograph).
3) The thick \Lya\ forest at these redshifts complicate the detection and measurements of these troughs. 

In contrast, the EUV500 covers many excited transitions from abundant species that produce outflow troughs (see figures \ref{fig:EUV500lines} and \ref{fig:1051spectrum}). Such usable troughs (allowing for an \ne\ determination) are detected in most such outflows (in 12 out of the 13 studied here).
Furthermore, the \Lya\ contamination is much smaller in redshift 0.5-1.5 objects.

An alternative method for determining \ne\ is via a photoionization time-scale analysis \cite[this time-scale is inversely proportional to the \ne\ of the outflow; see][]{Arav15}.
However, this method is both resource intensive (many observational epochs are needed) and is model-dependent \cite[see the discussion in section 7.1 of][] {Arav18}

\subsubsection{Abundance Determinations}\label{subsubsec:Abundances}

There are two advantages for attempting to determine abundances using EUV500 data:\\
1. High-ionization outflow data, covering $\lambda_{\rm rest}>1050$~\AA, usually show troughs from only H, C, N and Si (and in rare cases P and S). In the EUV500 we detect troughs from H, N, O, Ne, Na, Mg, Al, Si, S, Ar, Ca and Fe, allowing for the abundances of many more elements to be determined.\\
2. In the $\lambda_{\rm rest}>1050$~\AA\ region, C and N show troughs from only one ion (Si has two),
making an abundance determination highly uncertain. The standard method is to vary each abundance until all the $N_{\rm ion}$ curves yield the optimal solution on the \Nh, \Uh\ parameter space.  However this is more akin to assuming a set of abundances without any way to verify it. In the EUV500 we have several cases of troughs from different ions of the same element, for example
\oiii, iv, v, and vi, \neiv, v, vi, and viii.  Such occurrences yield abundance-independent \Uh, \Nh\ solutions (as they arise from the same element). The reliability of these solutions is higher as we do not have abundances as free parameters.  Once we have confidence in the elemental solutions,  making the solutions for different elements consistent would yield a better constrained set of abundances.  
In a future paper we will use these attributes to constrain the abundances of these EUV500 outflows.

\subsection{Comparison with X-Ray Observations of Seyfert and Quasar Outflows }\label{subsec:xray}
Spectroscopic X-ray observations of Seyfert outflows detect the so-called warm-absorbers.
In all cases, the need for two or more ionization phases is evident \citep[e.g.,][]{Netzer03,Steenbrugge05,Holczer07,Kaastra14,Behar17}. This is mainly due to the much larger spread in
IP of the observed ionic species [(e.g. \ov\ (IP = 114 eV) to \sxvi\
(IP = 3500 eV)]. Unfortunately, in most cases the X-ray spectra
lack the resolution to kinematically associate the warm absorber
with the UV absorber seen in the same object.

The EUV500 data has both the S/N and spectral resolution to kinematically associate troughs from the VHP with troughs from the HP. Therefore, we are confident that both phases come from the same $R$, which validate our calculation methods for $\dot{M}$ and $\dot{E_{k}}$. For elaborations of these issues, see the discussion in section 8.1 of \cite{Arav13}.

\subsection{Comparison with Earlier EUV500 Observations of Quasar Outflows}\label{subsec:earlier-obs}

Observations of EUV500 quasar outflows predate HST. \citet{Pettini86}  
observed BALQSO PG 0946+301 using the IUE satellite and identified EUV500 troughs from \niv, \oiv, \ov\ and  \neviii.  Soon after that, HST was launched with 2 orders of magnitude higher sensitivity in the UV. \cite{Korista92} used FOS to observe BALQSO 0226-1024. They measured $N_{\rm ion}$  for 14 ionic species from EUV500 troughs, but the measurements suffered from heavy trough blending as the full width of the outflow was 25,000 \kms. \cite{Telfer98}
observed the high-redshift BALQSO SBS 1542+541  using FOS, detected \sixii\ troughs, and made the first attempt to find the ionization solution using troughs from VHI species. \cite{Arav01} published a similar ionization solution using STIS observations of PG 0946+301.

In hindsight, all of these early efforts suffered from similar limitations:\\
1. The targeted outflows had trough widths larger than 4000 \kms\ (partially chosen because of the low spectral resolution of the spectrographs and the intent of observing BALs) and, therefore, suffered from considerable blending of troughs in the EUV500.\\
2. The resultant large uncertainties  in the measured $N_{\rm ion}$ did not allow for the identification of two different ionization phases. \\
3. Troughs from excited ionic states were either not identified or too blended to measuring their $N_{\rm ion}$ \cite[e.g., the \oiii* troughs detected in FOS observation of PG 0946+301]{Arav99b}. Therefore, no constraints could be put on the distance of the outflows from the central source ($R$).

COS enabled a significant breakthrough. \cite{Arav13} analyzed two outflows, seen in COS data, of quasar HE~0238-1904. The width of both outflows was around 600 \kms, enabling the measurements of absorption troughs from \oiv\ and \oiv* that lead to an $R$ determination. Accurate measurements of $N_{\rm ion}$
from ions with a wide range of ionization potential (\oiv\ to \mgx) revealed the necessity of a two-phase photoionization solution similar to the one depicted in figure \ref{fig:1051UHNH}. The combination of the large \Nh\ of the VHP, the high velocity and a large $R$, yielded a large value of $\dot{E_{k}}$ (see table \ref{tb:ParaSystems}). \cite{Finn14} used COS data of a narrow 
(full $\Delta v\sim600$ \kms), low-velocity outflow detected in quasar
FBQS J0209-0438. They determined an $R$ of 2000--6000 parsecs for different components of the outflow and showed the existence of two ionization phases.

Compared with the  results of \cite{Arav13},  the work presented here advances  our understanding of quasar outflows in several dimensions.\\
1. We show results for 13 outflows compared with the two analyzed in \cite{Arav13}. Thus, a) uncovering a wider outflow phenomenology, b) enabling comparative analysis, and c) putting on firmer ground the extrapolations of these results to quasar outflows in general.  For example, the conclusion that most high-ionization outflows observed from the ground have a dominant VHP that is revealed by EUV500 observations (see section \ref{sec:VHP}).  \\
2. The higher redshift of our targets let us measure troughs at shorter rest-frame wavelength, which increase the robustness of our photoionization analysis, and especially the ability to probe the VHP of the outflows. \\
3. We have several cases of more than one \ne\ diagnostic for a given outflow
(e.g., see figure 3 in Paper III). 
The availability of two or more diagnostics minimize the probability of systematic issues that can lead to erroneous $R$ determinations.

 \subsection{BAL Definition for the EUV500}
\label{sec:BAL}
\cite{Weymann91} defined the criteria for a quasar absorption trough being classified as a BAL. The definition requires a continuous \civ\ absorption over $\Delta v\geqslant$ 2000 km s$^{-1}$  at  $I \leqslant$ 0.9 (where $I$ is the normalized residual intensity) starting from at least --3000 km s$^{-1}$ blueward of the emission line center. The BAL definition is focused on \civ\  since it usually produces the strongest absorption troughs observed long-ward of the \Lya\ forest. There is a 500 km s$^{-1}$ velocity separation between the \civ\ doublets at 1548.19\angstrom\ and 1550.77\angstrom. Therefore, the BAL requirement for  a \civ\ absorption trough from a single transition is $\Delta v\gtrsim$ 1500 km s$^{-1}$. 

 \civ~absorption troughs are not covered in the EUV500. 
Therefore, we  define a BAL in the EUV500 region as: continuous absorption with $I \leqslant$ 0.9 over $\Delta v\gtrsim$ 1500 km s$^{-1}$, starting from at least --3000 km s$^{-1}$ blueward of the emission line center, seen in the widest, uncontaminated absorption trough (in our sample, it is either \neviii~770.41 \AA, \neviii~780.32 \AA\ or \ov~629.73 \AA). Mini-BALs \citep{Hamann04} are defined similarly with  1500 $\gtrsim \Delta v \gtrsim$ 500 km s$^{-1}$.  We note that 
the velocity separation between the \neviii\ doublet transitions is 3800 km s$^{-1}$. Therefore, if their absorption is self-blended, the outflow is clearly classified as a BAL.

\section{SUMMARY}
\label{sec:SUMMARY}

We executed the first 
dedicated HST/COS survey of quasar outflows covering the diagnostic-rich 500--1050~\AA\
rest-frame (hereafter, EUV500) spectral region.
This paper (Paper I) summarizes the main results of the survey 
and discuss their importance to various aspects of quasar research.
A detailed analysis of the data is presented in Papers II--VII (see section \ref{sec:Intro}).

Using 1 of the 13 outflows discussed here, we give an overview of how the scientific results are extracted from the data (see section~\ref{sec:path}), including: the photoionization solution, number density, distance from the central source ($R$) and kinetic luminosity ($\dot{E}_k$). 

In section \ref{sec:results} we give a summary of the results from these investigations,  
including: \\
1)	Measurements of the three most energetic outflows to date ($\dot{E}_k\geq10^{46}$ erg s$^{-1}$), which can be the main agents for AGN feedback processes in the environments of the host galaxies. \\
2)	All the outflows have a very high-ionization component, similar to the one found in warm absorbers, which carries most of their kinetic luminosity.  This detection suggests that all the high-ionization outflows detected from the ground also have a similar very high-ionization component. \\
3)	Of the 13 studied EUV500 outflows, 9 have $100<R<2000$ parsecs, 2 have $5<R<20$ parsecs, 1 has $0.05<R<50$ and in 1 cases, $R$ cannot be determined. \\
4)	One of the outflows has the largest velocity shift (1550 \kms) and acceleration (1.5 cm s$^{-2}$) measured to date. This outflow is physically similar to the fast X-ray outflow detected in quasar PG 1211+143. \\
5) These findings were partially enabled by the first detection of absorption troughs from previously unseen transitions of \oiv*, \ov*, \arvi, \nev*, \nevi*, \caiv, \cav, \cav*, \cavi, \cavii, \cavii*, \caviii\ and \caviii*.

We discuss the many advantages of studying quasar outflows using EUV500 data compared to ground-based observations, including: the ability to measure the dominant very high-ionization phase of the outflow, the ease and profound determination of the total \Nh. and the ionization structure of the outflows. For the majority of EUV500 outflows, $R$ can be determined, whereas it can be done in only a tiny fraction of ground-based spectra.

The VHP ionization parameter is similar to the lower ionization portion of the warm absorbers, which is seen in spectroscopic X-ray observations of Seyfert outflows. However, the warm absorbers reveal higher ionization phases through the detection of ions with much higher ionization potential. It is probable that such phases occur also in quasar outflows but the EUV500 does not cover the needed spectral diagnostics.

\acknowledgments
N.A., X.X.,  and T.M. acknowledge support from NSF grant AST 1413319, as well
as NASA STScI grants GO 14777, 14242, 14054, 14176, AR-15786, and NASA ADAP 48020.
G.K. acknowledge support from  NASA STScI grants GO 14777, 14242, 14054, and 14176; and NASA ADAP 48020.

Based on observations made with the NASA/ESA \textit{Hubble Space Telescope}, and obtained from the data archive at the Space Telescope Science Institute. STScI is operated by the Association of Universities for Research in Astronomy, Inc. under NASA contract NAS5-26555. CHIANTI is a collaborative project involving George Mason University, the University of Michigan (USA), University of Cambridge (UK), and NASA Goddard Space Flight Center (USA).

\end{document}